\documentclass[pra,twocolumn,showpacs]{revtex4}
\usepackage{amsmath,bbm,epsfig}
\usepackage{amsfonts}
\usepackage{mathrsfs}
\newcommand{\bra}[1]{\langle #1 | \,}

\newcommand{\ii}{\;\! \mbox{i} \;\!}

\newcommand{\mb}[1]{{\rm #1}}

\begin{document}

\title{Nonlinearity induced destruction of resonant tunneling
in the Wannier-Stark problem}

\author{S. Wimberger$^{1}$, R. Mannella$^{1}$, O. Morsch$^{1}$,
E. Arimondo$^{1}$, A.R. Kolovsky$^{2,3}$, and A. Buchleitner$^2$}
\affiliation{
$^1$Dipartimento di Fisica E. Fermi, CNR-INFM,
Unversit\`a di Pisa, Largo Pontecorvo 3, 56127 Pisa, Italy \\
$^2$Max-Planck-Institut f\"ur Physik komplexer Systeme,
N\"othnitzer Str. 38, 01187 Dresden, Germany \\
$^3$Kirensky Institute of Physics, 660036 Krasnoyarsk, Russia}

\date{\today}

\begin{abstract}
We present detailed numerical results on the dynamics of a Bose-Einstein
condensate in a tilted periodic optical lattice over many Bloch periods.
We show that an increasing atom-atom interaction systematically 
affects coherent tunneling, and eventually destroys the 
{\em resonant} tunneling peaks.
\end{abstract}


\maketitle

Experiments with cold and ultracold atoms made it possible in the last
decade to prepare and control the centre-of-mass motion of atoms
with unprecedented precision. Many toy models of either many-body
solid state physics \cite{BOfirst,BECBOfirst,pisaBO,nist,ingu}
or of simple Hamiltonian systems,
whose complexity arises from an external driving force \cite{driven},
were realized with the exceptional control offered by static
or time-dependent optical potentials.

Particularly Bose-Einstein condensates (BEC) whose
initial momentum spread can be adjusted in width
and absolute position have proved to be an extremely helpful experimental tool
\cite{pisaBO,nist,fallani,pisaASY,instabil}.
In addition, a BEC offers interesting new features originating
from the intrinsic interactions between the atoms.
Examples of such effects are new quantum phases
\cite{greiner}, soliton-like motion
\cite{eiermann}, the occurrence of energetic or dynamical instabilities
in condensates \cite{fallani,instabil,wuniu,zheng}, or the decay and
subsequent revival of Bloch oscillations (BO) \cite{revival}.

We focus on the evolution of a BEC
loaded into a one-dimensional lattice and subjected to an additional
static force $F$, which is most easily realized and controlled
by accelerating the optical lattice \cite{BOfirst,pisaBO}.
In previous experiments, a BEC was accelerated to allow for a single
crossing of the Brillouin zone (BZ), and two effects were
observed: for large accelerations,
an enhanced tunneling probability
from the ground state band to the first excited band due to the
atom-atom interaction was measured \cite{pisaBO,pisaASY}.
Secondly, for smaller accelerations (where tunneling is negligible)
signatures of a dynamical instability in the BEC were observed
\cite{instabil,footnote}. By contrast, here we investigate
the dynamics of a BEC performing many Bloch oscillations (BO), and
we ask ourselves how the atom-atom interaction
affects tunneling for a {\em sequence} of BZ crossings.
In particular, we scan $F$
to study the impact of the atom-atom interaction on
resonantly enhanced tunneling (RET),
for which the standard Landau-Zener prediction is modified
even in the absence of interactions \cite{kolo}.
The RET leads to a faster decay of the Wannier states trapped in the 
potential wells. 
With the survival probability and the recurrence probability (see Eqs.
(\ref{eq:sur}) and (\ref{eq:rec}) below) we present two consistent measures
for the nonlinear RET which define clear experimental signatures of the 
destruction of the coherent tunneling process inside the periodic potential.

If we neglect interactions for a moment, our
system will be described by the Hamiltonian
\begin{equation}
H = -\frac{\hbar^2}{2M}\frac{d^2}{dx^2} +
V \sin^2\left(\frac{\pi x}{d_L}\right) + Fx \;.
\label{eq:bloch}
\end{equation}
Here $d_L$ is the spatial period of the optical lattice
with maximal amplitude $V$, and $M$ the atomic
mass. Eq.~(\ref{eq:bloch}) defines the well-known Wannier-Stark problem,
which gives rise to BO with period
$T_{\rm Bloch}=h/d_LF$ ($h$ is Planck's constant).
If tunneling is small, we can view the system as moving at a constant speed
in momentum space within the fundamental BZ. At the zone
edge, most of the wave packet is reflected (giving rise to BO)
while a small part can tunnel across the first band gap
to the next higher-lying energy band and then escape quickly by successive
tunneling events across the smaller (higher) band gaps.
Landau-Zener theory predicts a decay rate \cite{kolo}
\begin{equation}
\Gamma (F) \propto F e^{-\frac bF}\,,
\label{eq:exp}
\end{equation}
where $b$ is proportional to the square of the energy gaps.
Eq.~(\ref{eq:exp}) is modified by RET which occurs
when two Wannier-Stark levels in neighboring potential wells
are coupled strongly due to their accidental degeneracy.
The RET results in pronounced
peaks in the tunneling rates, e.g., as a function of $1/F$,
on top of the global exponential decay described by (\ref{eq:exp})
\cite{kolo}. In this paper we investigate the impact of the effective
shift of the Wannier-Stark levels by a nonlinear interaction term.

For the linear problem (\ref{eq:bloch}), the decay rates have been measured
previously in the regime of short life-times in the ground state band
(of the order of 100 $\mu$s), where $\Gamma (F)$ 
is essentially smooth \cite{Raizen1997}.
Since RET is a coherent quantum effect, the peaks should be 
sensitively affected by the atom-atom interaction, 
which can be varied experimentally by changing either the density 
of the BEC or through the atom-atom scattering potential 
via a Fesh\-bach re\-so\-nance \cite{wieman2001}.
Our results are a consequence of {\em many} sequential Landau-Zener
events, and they show the destruction of a RET peak with increasing 
interaction strength, in a regime which is experimentally accessible.

We use a fully 3D Gross-Pitaevskii equation (GPE) 
\cite{Mannella1998} to describe the temporal evolution of a BEC
which is subject to realistic potentials:
\begin{eqnarray}
&\ii \hbar \frac{\partial }{\partial t}\psi (\vec{r},t) =
\left[-\frac{\hbar^2}{2M}\nabla ^2 + \frac{1}{2}M \left( \omega_x^2 x^2 +
\omega_{\rm r}^2 \rho^2 \right) + \right.
\nonumber \\
& \left.V \sin^2\left(\frac{\pi x}{d_L}\right)
+ F x + g N \left| \psi(\vec{r},t) \right|^2
\right] \psi(\vec{r},t) \;.
\label{eq:GP}
\end{eqnarray}
$\psi(\vec{r},t)$ represents the condensate
wave function, and the frequencies $\omega_x$
and $\omega_{\rm r}$ characterize the longitudinal and transverse
harmonic confinement (here with cylindrical symmetry: $\rho =
\sqrt{y^2 + z^2}$). We fixed $d_L=1.56 \;\rm \mu m$ and
$V/E_{\rm R}=5$ for our computations, with the
recoil energy $E_{\mb{R}}=p_{\mb{R}}^2/2M$ for
$p_{\mb{R}}=\hbar\pi/d_L$, and the  recoil
period $T_{\mb R}=h/E_{\mb R}$.
The above values for $d_L$ and $V$ were realized in the experiments
reported in \cite{pisaBO,pisaASY,instabil} based on two laser beams
propagating at an angle different from $\pi$. In Eq.~(\ref{eq:GP}),
the nonlinear coupling constant is given by $g=4\pi \hbar^2 a_s/M$, where
$a_s$ is the $s$-wave scattering length and $N$  the
number of atoms in the BEC \cite{O1998,Mannella1998}. The dimensionless
nonlinearity $C=gn_0/(8E_{\rm R})$ is computed from the peak density of the
initial state of the condensate,
with $C=0.027\ldots 0.31$ for the experimentally investigated
range of \cite{pisaBO}, and with $C=0.5$ reached in \cite{chu}.
Here we focus on $C>0$, but report briefly also 
on attractive interactions with $C<0$. 
The latter case leads to a fundamentally different behavior of the system
because the collapse of the condensate introduces an additional
time scale, which for experimentally relevant parameters is of the order
of 10 msec \cite{WHS2005,wieman2001} (slightly longer than 
$T_{\rm Bloch}=1.8\ldots3.0 \ \rm msec$ here). 

The GPE (\ref{eq:GP}) is numerically integrated using finite difference
propagation, adapted by a predictor-corrector estimate to reliably
evaluate the nonlinear interaction \cite{Mannella1998}.
Since our system is essentially the problem of a constantly
accelerated particle for the part of the wave function
which has tunnelled out of the first BZ already, one must
be careful with the application of absorbing boundary conditions 
or complex coordinate methods \cite{GPabs,SP2004}. 
To avoid any spurious effects due to the fast spreading,
we use a large numerical basis. In this way, we fully cover
the 3D expansion of the entire wave packet, including its tunnelled tail,
without the use of non-Hermitian potentials.
The initial state propagated by (\ref{eq:GP})
is the relaxed condensate wave function, adiabatically loaded into the
confining potential given by the harmonic trap and the
optical lattice (with $F=0$). Approximate analytic forms of the relaxed state
are found, e.g., in \cite{pedri}, but we used an imaginary time
propagation to reliably compute the initial state for $C>0$.

\begin{figure}
\centerline{\epsfig{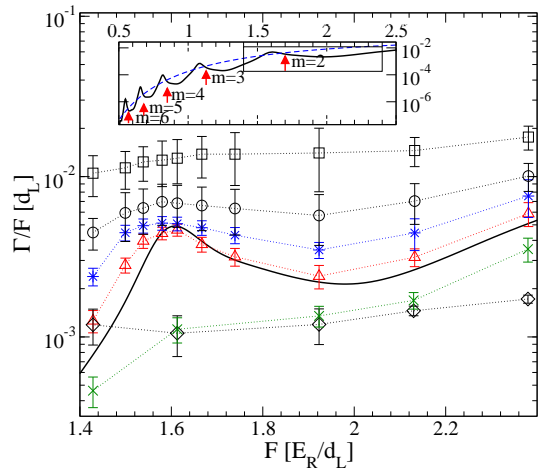}}
\caption{(color online).
Tunneling rates obtained by exponential fits to data of
$P_{\rm sur}(t)$ as the solid-line fits in Fig. \ref{fig:3}.
Here the peak in the box in the inset is scanned locally,
while globally the rates follow an exponential law (dashed line in the inset).
$C=-0.31$ (diamonds), $C=-0.065$ (crosses), $C=0$ (solid line), 
$C=0.027$ (pyramids), $C=0.065$ (stars), $C=0.12$ (circles), 
$C=0.31$ (squares).
The ``error'' bars interpolate different exponential fits such as the
dot-dashed ones in Fig. \ref{fig:3}. The arrows in the inset mark the
peak positions as predicted by the simple argument stated in the text.
}
\label{fig:1}
\end{figure}

The linear decay rates for non-interacting
atoms in the optical lattice are
computed from the spectrum of the 1D Wannier-Stark problem of
Eq.~(\ref{eq:bloch}) using, e.g., the method of \cite{kolo}.
Those linear rates are plotted in Fig.~\ref{fig:1}.
The maxima in the rates occur when $Fd_Lm$ (with $m$ integer)
is close to the difference between the first two energy bands
(averaged over the BZ) of the $F=0$ problem \cite{kolo}.
The actual peaks are slightly shifted with respect to the above
estimate (marked by arrows in the inset of Fig.~\ref{fig:1}),
owing to a field-induced level shift \cite{kolo}.

Experimentally, the most easily measurable quantity is the momentum
distribution of the BEC obtained from a free expansion 
after the evolution inside the lattice.
From the momentum distribution we determine the survival probability
by projection of the evolved state $\psi(\vec{p},t)$ onto the support
of the initial state
\begin{equation}
P_{\rm sur}(t) \equiv \int_{-p_{\rm c}}^{p_{\rm c}} dp_x
\left( \int dp_ydp_z |\psi(\vec{p},t)|^2 \right)\;,
\label{eq:sur}
\end{equation}
where $p_{\rm c} \geq  3p_{\mb R}$  is a good choice 
since three momentum peaks are initially significantly populated, 
corresponding to $-2p_{\mb R},0,2p_{\mb R}$ \cite{pisaBO,pedri}.

\begin{figure}
\centerline{\epsfig{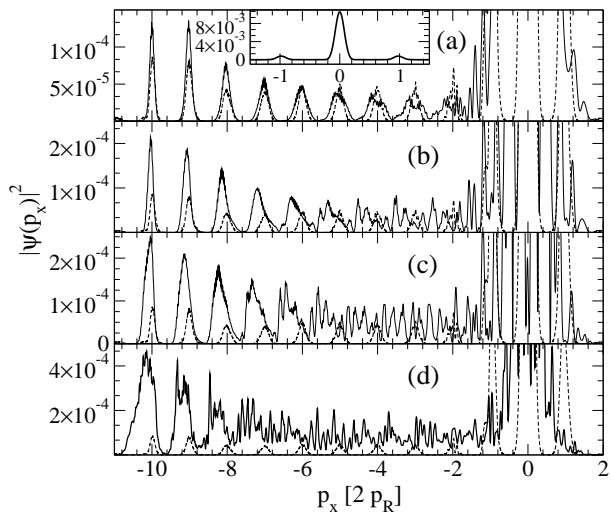}}
\caption{Momentum distributions after 10 BOs,
for $T_{\rm R}/T_{\rm Bloch}=1.428$. $C=0$ (dotted) compared with (a)
$C=0.027$ (full line; the inset shows the corresponding 
$t=0$ distribution), (b) $C=0.065$, (c) $C=0.12$, and (d) $C=0.31$.
}
\label{fig:2}
\end{figure}

Figure \ref{fig:2} shows the initial population in momentum
space (inset in (a)) as compared with the population after $10$ BO periods,
for both the linear and the nonlinear case. 
The increase of $C>0$ has two effects:
firstly, it enhances the tunneling for the first few crossings
of the BZ.
Secondly, it scrambles the out-coupled part of the wave function
(see Fig. \ref{fig:2} and its complement in Fig.
\ref{fig:4} below), as previously observed in \cite{pisaBO,ingu,BECBOfirst}.
The change in the momentum distributions after various Landau-Zener
events is a manifestation of the intrinsic instability of the nonlinear
GPE dynamics \cite{instabil,wuniu}.

\begin{figure}
\centerline{\epsfig{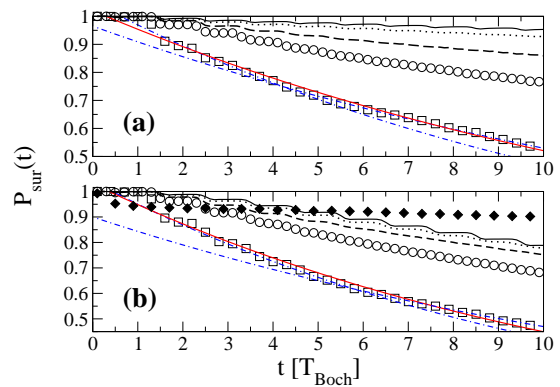}}
\caption{(color online).
$P_{\rm sur}(t)$
for (a) $T_{\rm R}/T_{\rm Bloch}=1.428$ and (b) $1.613$ (the
peak maximum in Fig. \ref{fig:1}).
$C$ is scanned from $-0.31$ (diamonds in (b) only),
$0$ (solid line), $0.027$ (dotted line),
$0.065$ (dashed line), $0.12$ (circles) to $0.31$ (squares).
The gray/red solid lines show global exponential fits to the $C=0.31$ data,
while the dot-dashed lines show exponential fits for small
and large $t$ respectively. From those fits, the rates in
Fig.~\ref{fig:1} and their systematical variation in time are obtained.
The step-like structures reflect the periodic BO
and are correlated with the dephased oscillations in Fig.~\ref{fig:4}.
}
\label{fig:3}
\end{figure}

Instead of studying the details of the distributions shown in
Fig.~\ref{fig:2}, we will focus on the temporal decay of the
survival probability in the following. Figure ~\ref{fig:3} presents
$P_{\rm sur}(t)$, which for the linear case has an exponential form 
(apart from the $t \to 0$ limit \cite{raizenN})
\begin{equation}
P_{\rm sur}(t) \sim e^{- \frac{t\Gamma}{\hbar} }\;,
\label{eq:surexp}
\end{equation}
with the characteristic exponent $\Gamma$. The temporal
behavior of $P_{\rm sur}$ depends significantly on $C$.
For $C = \pm 0.31$, we observe clear deviations from a purely
exponential decay, as present for small $C$.
A {\em repulsive} nonlinearity initially enhances the tunneling
more than after about five crossings of the BZ (see fits to data
in Fig.~\ref{fig:3}). This deviation from the
mono-exponential behavior means that the tunneling events
occurring at different integer multiples of the Bloch period
are correlated by the presence of the nonlinearity.
Since the remaining density becomes smaller, the impact of the 
nonlinearity becomes less. The result is
that the rate $\Gamma$ is defined only locally in time, and its
value systematically decreases as time increases.

An {\em attractive} interaction can stabilize the system
at the RET peak, which is shown for $C=-0.31$ in Fig. 3(b).
For optimal comparison, we chose the same initial state
(for $C=+0.31$) which then was evolved for $F\neq 0$ with
$C=-0.31$. Such a scenario could be realized by a sudden change 
of the sign of the scattering length through a Feshbach resonance 
\cite{wieman2001}. This result is consistent with studies of 
simpler models, where a resonance state can be stabilized at
system-specific strengths of the nonlinearity \cite{SP2004,korsch2005}.

The impact of the nonlinearity on the dynamical evolution of the ``closed''
system confined to the fundamental BZ can
be studied with the help of the recurrence probability
\cite{Raizen1997}, defined by the autocorrelation
\begin{equation}
P_{\rm rec}(t) \equiv \left| \bra{\psi(t)}\psi(t=0)\rangle \right|^2\;.
\label{eq:rec}
\end{equation}
The BO manifest themselves as the periodic oscillations in
$P_{\rm rec}(t)$ plotted in Fig. \ref{fig:4}. These oscillations 
are less and less pronounced with increasing $C$, in much 
the same way as the momentum
peaks are washed out when the band edge is crossed in the regime
of instability \cite{instabil}. In contrast to the
survival probability, $P_{\rm rec}$ is a phase sensitive
measure, and therefore it shows -- in addition to the temporal decay --
the dephasing of the BO due to the nonlinearity.
For $C=0$, the recurrence maxima
decay in time with the same rate
as $P_{\rm sur}(t)$, which offers an alternative method for
extracting $\Gamma$. For $C\neq 0$, $P_{\rm rec}$
can be integrated over time, and the
rates are extractable by the approximate proportionality between the
integrated area and the inverse decay rate (recalling that
$\int dt f(t)\exp(-t\Gamma) \sim 1/\Gamma$ to leading order,
for a periodic function $f(t)$). The latter approach
works because we can determine the linear rate from a direct fit
to $P_{\rm rec}$ and then compare the ratio of the linear and
the nonlinear area (denoted by $A_0$ and $A_C$). This
rough estimate $\Gamma_C \sim \Gamma_0 A_0/A_C$
agrees within $25\%$ with the rate extracted from the fits
to the data of Fig. \ref{fig:3}. The estimate could be improved
if we knew the analytic form of the function $f(t)$, and it breaks
down for large $C$, when the periodic oscillations in $P_{\rm rec}$ are
destroyed.

\begin{figure}
\centerline{\epsfig{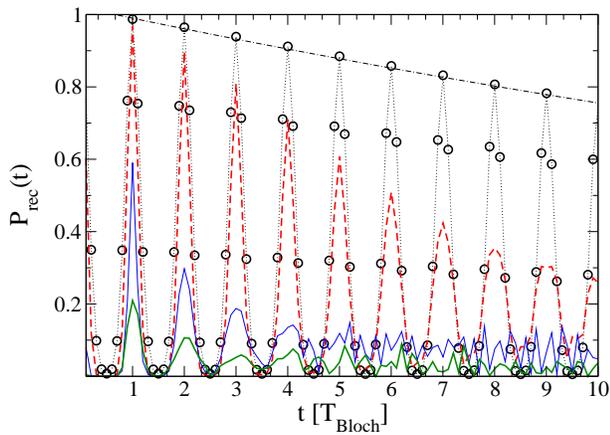}}
\caption{(color online).
$P_{\rm rec}(t)$ for the data shown
in Fig. \ref{fig:3}(b) with $C=0$ (circles), $0.027$ (dashed), 
$0.12$ (thin line), and $0.31$ (thick line). The dot-dashed line
presents an exponential fit to the maxima of the $C=0$ data.
}
\label{fig:4}
\end{figure}

Having introduced two methods to extract the tunneling rates,
we scan the parameter $F$ across a RET  peak of the globally exponential
curve $\Gamma(1/F)$ (see Fig.~\ref{fig:1}).
The scanned range in $F$ corresponds
to values of lattice accelerations between $0.99\; \rm ms^{-2}$ and
$1.65 \; \rm ms^{-2}$, which are standard in experiments
\cite{pisaBO,instabil}.

A repulsive nonlinearity particularly affects
the wings of the peak and, for small $C$, much less the peak maximum.
The global increase of $\Gamma$ with increasing $C$
is qualitatively predicted in \cite{niu}, with
enhanced {\em single} Landau-Zener crossing probabilities
induced by the effective reduction of the energy gap due to the nonlinearity.
The left and right-most points in Fig.~\ref{fig:1}
are in the regime where an amended version of (\ref{eq:exp}) indeed
applies \cite{niu}, and here $\Gamma/F$ is approximately proportional to $C$.
However, near the peak, the rates do not follow a simple scaling
law as a function of $C$, and the argumentation of \cite{niu} does not apply.

For $C<0$ we also observe the destruction of the RET peak. For
$C=-0.065$, the BEC clearly stabilizes in the potential wells, whilst
for $C=-0.31$ the situation is more complicated (see Fig.~\ref{fig:1}).
The precise dynamics of the system is governed by the two 
separate time scales for tunneling and collapse, which strongly depend 
on parameters in the sensitive RET regime.

In an experiment, $\omega_x$ can either be set to zero or decreased
to $\omega_x/2\pi \lesssim 1\;\rm Hz$ to realize
a quasi 1D {\em nonlinear} Wannier-Stark problem. We verified that
letting $\omega_x$ tend to zero for the evolution
with $F \neq 0$, or applying a small finite $\omega_x$
gives the same results for the BO cycles studied here.
Furthermore, for $0 < C \lesssim 0.05$,
using the renormalized nonlinearity of \cite{O1998} we 
observed that a 1D version of Eq.~(\ref{eq:GP})
reproduces well the 3D data.
If $|C|$ is larger, the nonlinearity couples the
longitudinal and transverse degrees of freedom, which affects
the dynamics of a real BEC in a
non-trivial way \cite{Mannella1998}.
The 1D computations are feasible up to 100 Bloch periods,
and this would allow one to extract the tunneling rates more reliably. The
effect of the nonlinearity is, however, hardly visible for $0<C<0.05$,
and quantitative predictions for a broad range of $C$ relied on
3D computations.

To summarize, we observed and quantified the deformation and
destruction of the RET peaks due to interactions in
a BEC in an accelerated optical lattice. Our results
complement ongoing studies of interaction-induced
processes such as dynamical instabilities or the decay and 
subsequent revival of BO. In the regime of small nonlinearity,
where dynamical instabilities are not fully developed, the survival
and recurrence probabilities experience an exponential decay
modified by the condensate nonlinearity. The temporal decay of these
observables remains a useful indicator also for large nonlinearity, even
if the resonant structure in the tunneling rate is washed out.

We thank M. Cristiani and D. Ciampini for helpful
discussions and the Humboldt Foundation (Feodor-Lynen Program),
MIUR COFIN-2004, and ESF (QUDEDIS) for support.


\begin{thebibliography}{20}

\bibitem{BOfirst}
M. BenDahan {\em et al.},
Phys. Rev. Lett. {\bf 76}, 4508 (1996);
S.R. Wilkinson {\em et al.},
{\em ibid.} {\bf 76}, 4512 (1996).

\bibitem{BECBOfirst}
B.P. Anderson and M.A. Kasevich,
Science {\bf 282}, 1686 (1998).

\bibitem{pisaBO}
O. Morsch {\em et al.},
Phys. Rev. Lett. {\bf 87}, 140402 (2001);
M. Cristiani {\em et al.},
Phys. Rev. A {\bf 65}, 063612 (2002).

\bibitem{nist}
J. Hecker-Denschlag {\em et al.},
J. Phys. B {\bf 35}, 3095 (2002).

\bibitem{ingu}
G. Roati {\em et al.},
Phys. Rev. Lett. {\bf 92}, 230402 (2004).

\bibitem{driven}
M.G. Raizen, Adv. At. Mol. Opt. Phys. {\bf 41}, 43 (1999);
D.A. Steck, W.H. Oskay, and M.G. Raizen, Science 293, 274 (2001);
W.K. Hensinger {\em et al.}, Nature (London) 412, 52 (2001).

\bibitem{fallani}
L. Fallani {\em et al.}, Phys. Rev. Lett. {\bf 93}, 140406 (2004).

\bibitem{pisaASY}
M. Jona-Lasinio {\em et al.},
Phys. Rev. Lett. {\bf 91}, 230406 (2003).

\bibitem{instabil}
M. Cristiani {\em et al.}, Opt. Express {\bf 12}, 4 (2004).

\bibitem{greiner}
M. Greiner {\em et al.},
Nature (London) {\bf 415}, 39 (2002).

\bibitem{eiermann}
B. Eiermann {\em et al.},
Phys. Rev. Lett. {\bf 92}, 230401 (2004).

\bibitem{wuniu}
B. Wu and Q. Niu, New J. Phys. {\bf 5}, 104 (2003).

\bibitem{zheng}
Y. Zheng, M. Kostrun, and J. Javanainen,
Phys. Rev. Lett. {\bf 93}, 230401 (2004).

\bibitem{revival}
A.R. Kolovsky, Phys. Rev. Lett. {\bf 90}, 213002 (2003);
A. Buchleitner and A.R. Kolovsky,
{\em ibid.}, {\bf 91}, 253002 (2003);
Q. Thommen, J.C. Garreau, and V. Zehnle,
{\em ibid.}, {\bf 91} 210405  (2003).


\bibitem{footnote}
Instability at the band edge
has also been studied using a different experimental
protocol in \cite{fallani}.

\bibitem{kolo}
M. Gl\"uck, A.R. Kolovsky, and H.J. Korsch,
Phys. Rev. Lett. {\bf 83}, 891 (1999);
Phys. Rep. {\bf 366}, 103 (2002).

\bibitem{Raizen1997}
C.F. Bharucha {\em et al.},
Phys. Rev. A {\bf 55}, R857 (1997).

\bibitem{wieman2001}
J.L. Roberts {\em et al.},
Phys. Rev. Lett. {\bf 86}, 4211 (2001).

\bibitem{Mannella1998}
E. Cerboneschi {\em et al.},
Phys. Lett. A {\bf 249}, 495 (1998);
S. Wimberger {\em et al.},
Phys. Rev. Lett. {\bf 94}, 130404 (2005).

\bibitem{O1998}
M. Olshanii,
Phys. Rev. Lett. {\bf 81}, 938 (1998).

\bibitem{chu}
N. Gemelke {\em et al.}, 
cond-mat/0504311.

\bibitem{WHS2005}
S. W\"uster, I.J. Hope, and C.M. Savage,
Phys. Rev. A {\bf 71}, 033604 (2005).

\bibitem{GPabs}
T. Paul, K. Richter, and P. Schlagheck,
Phys. Rev. Lett. {\bf 94}, 020404 (2005).

\bibitem{SP2004}
P. Schlagheck and T. Paul, cond-mat/0402089;
N. Moiseyev and L.S. Cederbaum, cond-mat/0406189.

\bibitem{pedri}
P. Pedri {\em et al.},
Phys. Rev. Lett. {\bf 87}, 220401 (2001).

\bibitem{raizenN}
S.R. Wilkinson {\em et al.},
Nature (London) {\bf 387}, 575 (1997).

\bibitem{korsch2005}
D. Witthaut, S. Mossmann, and H.J. Korsch,
J. Phys. A {\bf 38}, 1777 (2005).

\bibitem{niu}
D.I. Choi and Q. Niu,
Phys. Rev. Lett. {\bf 82}, 2022 (1999);
O. Zobay, B.M. Garraway,
Phys. Rev. A {\bf 61}, 033603 (2000).

\end{thebibliography}
\end{document}